\documentclass[conference]{IEEEtran}
\IEEEoverridecommandlockouts

\usepackage{cite}
\usepackage{amsmath,amssymb,amsfonts}
\usepackage{algorithmic}
\usepackage{graphicx}
\usepackage{textcomp}
\usepackage{xcolor}

\usepackage{booktabs}
\usepackage{subfig}
\usepackage{amsmath,amsfonts}
\usepackage{algorithmic}
\usepackage{algorithm}
\usepackage{array}
\usepackage{textcomp}
\usepackage{stfloats}
\usepackage{url}
\usepackage{verbatim}
\usepackage{graphicx}
\usepackage{xcolor}
\usepackage{enumitem}
\usepackage{arydshln}
\usepackage{fancyhdr}
\usepackage{varwidth}
\usepackage{arydshln}
\usepackage{multirow}
\usepackage{mathtools}
\usepackage{datetime}
\usepackage{lipsum}
\usepackage{indentfirst}

\usepackage{kotex}
\usepackage{ragged2e}

\newcommand{\sys}{\textsf{RT-SNN}}

\newcommand{\npfp}{$\textsf{NPFP}^\textsf{min}$}
\newcommand{\npfpmem}{$\textsf{NPFP}^\textsf{mem}$}

\newcommand{\jl}[1]{{\color{black}{#1}}} 
\newcommand{\jll}[1]{{\color{black}{#1}}} 

\newcommand{\hb}[1]{{\color{black}{#1}}} 

\definecolor{caribbeangreen}{rgb}{0.0, 0.6, 0.6}
\newcommand{\kdh}[1]{{\color{black}{#1}}} 

\newcommand{\remove}[1]{}

\newtheorem{lemma}{Lemma}
\newtheorem{theorem}{Theorem}

\def\BibTeX{{\rm B\kern-.05em{\sc i\kern-.025em b}\kern-.08em
    T\kern-.1667em\lower.7ex\hbox{E}\kern-.125emX}}
    
\begin{document}

\title{Real-Time Scheduling Framework for Multi-Object Detection via Spiking Neural Networks
}

\author{
\IEEEauthorblockN{
Donghwa Kang\IEEEauthorrefmark{2},
Woojin Shin\IEEEauthorrefmark{3},
Cheol-Ho Hong\IEEEauthorrefmark{4},
Minsuk Koo\IEEEauthorrefmark{3},
Brent ByungHoon Kang\IEEEauthorrefmark{2},\\
Jinkyu Lee\IEEEauthorrefmark{5} and
Hyeongboo Baek\IEEEauthorrefmark{3}
}

\IEEEauthorblockA{\small{~}}

\IEEEauthorblockA{KAIST\IEEEauthorrefmark{2}}
\IEEEauthorblockA{University of Seoul\IEEEauthorrefmark{3}}
\IEEEauthorblockA{Chung-Ang University\IEEEauthorrefmark{4}}
\IEEEauthorblockA{Sungkyunkwan University\IEEEauthorrefmark{5}}

}

\maketitle

\begin{abstract}
\label{sec:abstract}
Given the energy constraints in autonomous mobile agents (AMAs), such as unmanned vehicles, spiking neural networks (SNNs) are increasingly favored as a more efficient alternative to traditional artificial neural networks. 
AMAs employ multi-object detection (MOD) from multiple cameras to identify nearby objects while ensuring two essential objectives, (\textsf{R1}) \textit{timing guarantee} and (\textsf{R2}) \textit{high accuracy} for safety.
In this paper, we propose \sys{}, the \textit{first} system design, aiming at achieving \textsf{R1} and \textsf{R2} in SNN-based MOD systems on AMAs. 
Leveraging the characteristic that SNNs gather feature data of input image termed as membrane potential, through iterative computation over multiple timesteps,
\sys{} provides multiple execution options with adjustable timesteps and a novel method for reusing membrane potential to support \textsf{R1}. 
Then, it captures how these execution strategies influence \textsf{R2} by introducing a novel notion of mean absolute error and membrane confidence. 
Further, \sys{} develops a new scheduling framework consisting of offline schedulability analysis for \textsf{R1} and a run-time scheduling algorithm for \textsf{R2} using the notion of membrane confidence.
We deployed \sys{} to Spiking-YOLO, the SNN-based MOD model derived from ANN-to-SNN conversion, and our experimental evaluation confirms its effectiveness in meeting the \textsf{R1} and \textsf{R2} requirements while providing significant energy efficiency.
\end{abstract}

\begin{IEEEkeywords}
Spiking neural network, autonomous mobile agents, object detection, timing guarantee
\end{IEEEkeywords}

\section{Introduction}
\label{sec:intro}

Autonomous mobile agents (AMAs), such as unmanned autonomous vehicles, rely on multi-object detection (MOD) using multiple cameras to identify and classify surrounding objects, like cars or pedestrians, by analyzing continuous input frames~\cite{KCK22, KLC22}. 
These perception results are crucial for time-sensitive operations like braking and must be delivered to the control subsystem within strict deadlines to ensure the safety of AMAs. 
Inaccurate perceptions could lead to misjudgments and potential accidents, so MOD subsystems in AMAs must provide (\textsf{R1}) \textit{timing guarantee} and (\textsf{R2}) \textit{high accuracy}.

Despite recent advancements in artificial neural networks (ANNs), their high power consumption limits their use in battery-dependent systems like AMAs~\cite{DSF21}.
To address energy constraints, spiking neural networks (SNNs) have emerged as an alternative~\cite{PRVS23, PVS23}. 
Unlike ANNs, which operate within a single timestep, SNNs process inputs over multiple timesteps, activating only during spike events. 
Neurons accumulate membrane potential over successive timesteps; once a threshold is reached, a binary output in a spike is triggered. 
This reduces computation and improves energy efficiency, especially on neuromorphic hardware optimized for SNN models.

\begin{figure}
    \centering
    \includegraphics[width=1\linewidth]{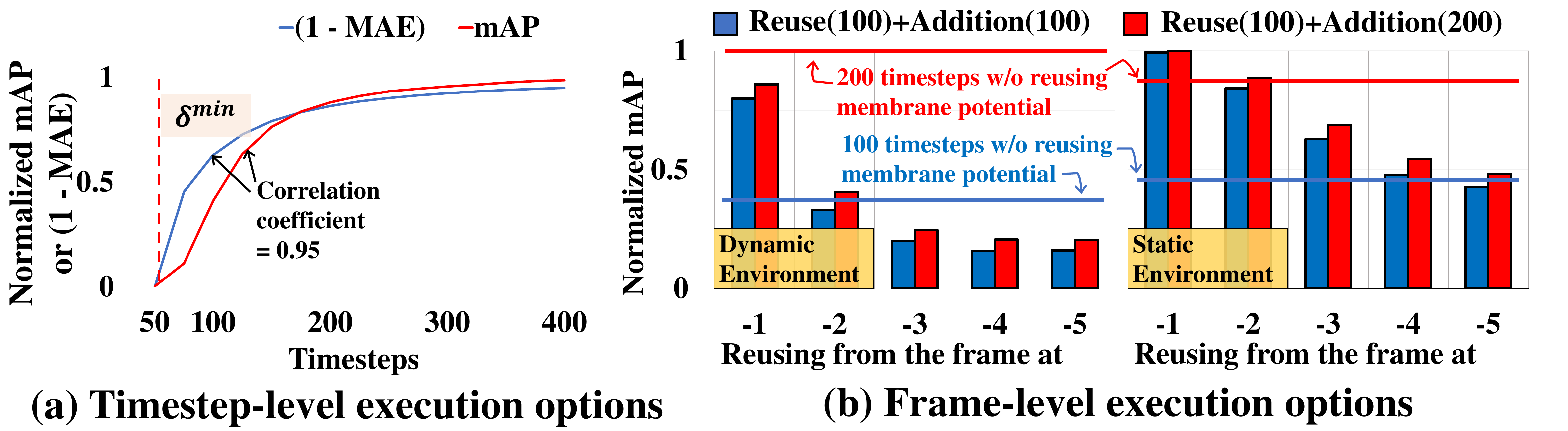}
    \caption{\hb{Influence of (a) timestep variations and (b) membrane potential reuse on accuracy}}
    \label{fig:mae_motivation_01}
    \vspace{-0.5cm}
\end{figure}

SNNs offer significant energy benefits (e.g., Spiking-YOLO on TrueNorth~\cite{MAA14} is 280 times more efficient than Tiny-YOLO on an NVIDIA Tesla V100 GPU~\cite{KPN20}), but integrating them into resource-limited AMAs presents challenges due to the trade-off between \textsf{R1} and \textsf{R2}. 
Due to the unique nature of SNNs, feature information (termed \textit{spike feature}~\cite{WSZ21}) is extracted progressively from input over multiple timesteps. 
As shown in Fig.~\ref{fig:mae_motivation_01}(a) with the KITTI dataset and Spiking-YOLO, detection accuracy (i.e., mAP) increases with more timesteps (indicating longer execution time), as noted in state-of-the-art SNN studies~\cite{KPN20, KPNKY20}. 
Thus, prioritizing fewer timesteps for \textsf{R1} may compromise \textsf{R2}, while optimizing for \textsf{R2} could jeopardize \textsf{R1}.

In this paper, we propose \sys{}, the first system design that achieves both \textsf{R1} and \textsf{R2} for SNN-based resource-limited MOD systems.
\sys{} provides multiple SNN execution options, each with different execution times for \textsf{R1}, and captures their impact on accuracy for \textsf{R2}.
\sys{} then develops a scheduling framework to identify the best execution option for every SNN execution, ensuring both \textsf{R1} and \textsf{R2}.
To do this, \sys{} first tackles the following challenge:
\begin{itemize}
    \item [\textbf{\textsf{C1}.}] How to provide multiple execution options for \textsf{R1}?
\end{itemize}

\noindent
To address \textsf{C1}, \sys{} offers \textit{timestep-level} execution options by dynamically adjusting timesteps for each MOD execution at time $t$ at runtime. 
Additionally, \sys{} allows the reuse of membrane potential extracted from the previous frame at $t$-$f$ (for $f\ge 1$), providing a \textit{frame-level} execution option (to be detailed in Sec.~\ref{sec:system_design}).

Each combination of timestep/frame-level execution options (i.e., given timesteps with or without membrane potential reuse (MPR)) affects \textsf{R2}, leading to the next challenge:

\begin{itemize}
    \item [\textbf{\textsf{C2.}}] How to capture the \textit{run-time} impact of timestep/frame-level execution options on \textsf{R2}?
\end{itemize}

\noindent
To address \textsf{C2}, we introduce the concept of \textit{membrane confidence}. 
For timesteps without MPR, we focus on \textit{mean absolute error} (MAE), which captures the disparity in spike features across consecutive timesteps. 
As shown in Fig.~\ref{fig:mae_motivation_01}(a), the inverse of normalized MAE closely matches normalized mAP (e.g., Pearson correlation coefficient is 0.95 in Fig.~\ref{fig:mae_motivation_01}(a)).
Using MAE, membrane confidence quantifies the expected accuracy of the membrane potential (to be detailed in Sec.~\ref{subsec:without}).

For timesteps with MPR, Fig.~\ref{fig:mae_motivation_01}(b) shows how reusing membrane potentials from 100 timesteps at $t$-$f$ (where $1 \le f \le 5$) affects accuracy at $t$, after 100 (blue bar) or 200 (red bar) additional timesteps, in both dynamic (left) and static (right) environments. Solid lines represent scenarios without membrane potential reuse.
As seen in Fig.~\ref{fig:mae_motivation_01}(b), using potentials from earlier than $t$-$2$ decreases mAP in dynamic environments, while even using $t$-$4$ potentials improves mAP in static settings. 
Over 200 timesteps at time $t$, not reusing membrane potentials (red bar) generally yields higher accuracy than reusing, except for $t$-1.

Based on this observation, membrane confidence quantifies (i) the expected accuracy (by MAE) of membrane potential extracted at $t$-$f$, (ii) its variance (by environmental dynamics) over $f$ frames, and (iii) the combined effect of \hb{(i)}, \hb{(ii)}, and additional potential extracted over given timesteps at $t$ (to be detailed in Sec.~\ref{subsec:with}).

The final challenge is to effectively utilize \textsf{C1} and \textsf{C2} to achieve \textsf{R1} and \textsf{R2}.

\begin{itemize}
    \item [\textbf{\textsf{C3.}}] How to use the answers to \textsf{C1} and \textsf{C2} to achieve \textsf{R1} and \textsf{R2}?
\end{itemize}

\noindent
To address \textsf{C3}, \sys{} develops a novel scheduling framework that assesses the feasibility of completing execution options (timesteps with or without MPR) within deadlines (\textsf{R1}) using \textit{schedulability analysis}, and finds the best feasible combination for maximum membrane confidence (\textsf{R2}) using a \textit{scheduling algorithm} (to be detailed in Sec.~\ref{sec:scheduling}).

\sys{} is the first to simultaneously address \textsf{R1} and \textsf{R2} for SNN-based MOD in resource-limited environments.
While prior works~\cite{KPN20, SCH23} advanced SNN-based MOD, they did not account for run-time accuracy at specific timesteps or timing guarantees. 
\sys{} is compatible with any SNN model for MOD, regardless of whether based on surrogate back-propagation or ANN-to-SNN conversion~\cite{KPN20}. 
Our evaluation on Spiking-YOLO~\cite{KPN20} with the KITTI dataset~\cite{APR12} shows that \sys{} outperforms existing methods in accuracy, energy consumption, and timing guarantees.

\section{\sys{}: Methodology}
\label{sec:system_design}

\begin{figure*}
    \centering
    \includegraphics[width=1\linewidth]{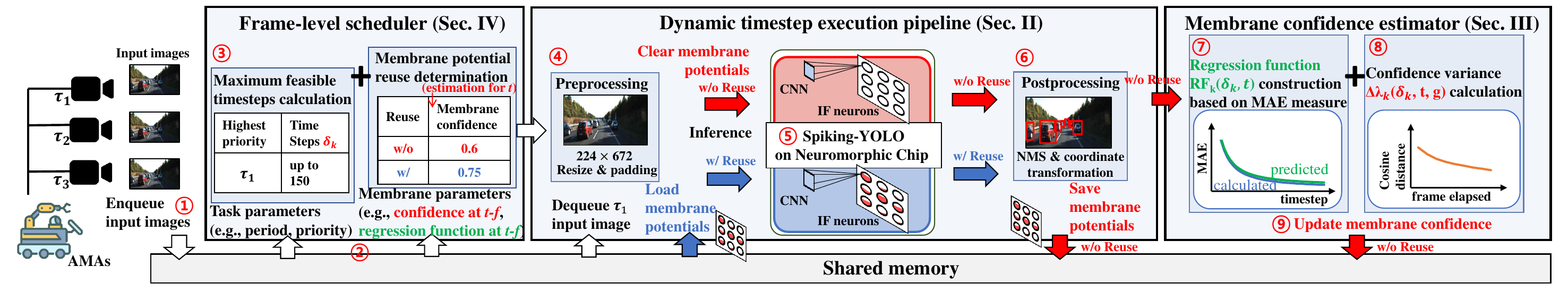}
    \caption{\sys{} Methodology}
    \label{fig:SNN_system_design}
    \vspace{-0.5cm}
\end{figure*}

\sys{} is designed to target a MOD system in AMAs equipped with $n$ cameras and share a single neuromorphic chip. 
Each camera on the AMAs may operate at the same or different frames per second (FPS), tailored to its functional importance (e.g., the front camera may operate at a higher FPS due to its critical role in navigation).
The components of \sys{}, which are \textit{the frame-level scheduler}, \textit{the dynamic timestep execution pipeline}, and \textit{the membrane confidence estimator}, 
are implemented as concurrent threads. 
These components utilize a shared memory architecture to facilitate efficient inter-thread communication. 
Fig.~\ref{fig:SNN_system_design} presents overall architecture of \sys{}.

\textbf{Frame-level scheduler.}
Initially, input images from the AMAs' multiple cameras are enqueued in a task queue within the shared memory (① in Fig.~\ref{fig:SNN_system_design}). 
For the $t$-th frame of the highest-priority task $\tau_k$, the frame-level scheduler retrieves the task parameters (e.g., period, priority) and membrane parameters (e.g., membrane confidence, regression function\footnote{The membrane confidence estimator develops this regression function using timestep and MAE pairs from the $(t$-$f)$-th frame to predict accuracy before executing the designated timesteps in the $t$-th frame, to be discussed in Sec.~\ref{subsec:without}.}) from the $(t$-$f$)-th frame (② in Fig.~\ref{fig:SNN_system_design}). 
The following are key features of the frame-level scheduler to be detailed in  Sec.~\ref{sec:scheduling}:

\begin{itemize} [leftmargin=*]
    \item \textit{Schedulability analysis}: the scheduler ascertains the maximum number of feasible timesteps $\delta_k$ that does not compromise timing guarantees.
    \item \textit{Scheduling algorithm}: the scheduler determines whether to reuse the membrane potentials from the $(t$-$f$)-th frame, which is based on a comparison of the estimated membrane confidence of w/o and w/ MPR (③ in Fig.~\ref{fig:SNN_system_design}). 
\end{itemize}

\textbf{Dynamic timestep execution pipeline.}
After the scheduling decision, the input image undergoes preprocessing, including resizing and padding to fit the dimensions required by the object detection algorithm (e.g., $224 \times 672$) (④ in Fig.~\ref{fig:SNN_system_design}). 
The processed image is then forwarded to the neuromorphic chip. 
The execution proceeds according to two scenarios, each determined by the scheduling decision.

\begin{itemize}[leftmargin=*]
    \item w/o MPR: the membrane potentials for all integrate-and-fire (IF) neurons are reset, and the model proceeds with SNN-based inference using the determined number of timesteps $\delta_k$ (⑤ in Fig.~\ref{fig:SNN_system_design}). The deployed SNN model's membrane potentials are preserved in the shared memory and the spike feature $s_k(\delta_k, t)$ extracted with $\delta_k$ at $t$ is recorded for use (to be detailed in Eq.~\eqref{eq:MAE}) in the membrane confidence estimator. 
    \item w/ MPR: the corresponding potentials from the $(t$-$f$)-th frame stored in the shared memory are retrieved and mapped onto the model prior to inference with the deployed SNN model (e.g., Spiking-YOLO) for the $t$-th frame. Then, the model conducts SNN-based inference with the determined number of timesteps $\delta_k$ (⑤ in Fig.~\ref{fig:SNN_system_design}).
\end{itemize}

\noindent
After inference, the results undergo further postprocessing steps, such as non-maximum suppression (NMS) and coordinate transformation (⑥ in Fig.~\ref{fig:SNN_system_design}).

\textbf{Membrane confidence estimator.}
This component is activated when membrane potentials from the $(t$-$f$)-th frame are not reused for the $t$-th frame, adhering to \sys{}'s operational policy (i.e., timesteps w/o MPR). 
The membrane confidence estimator has the following key features, to be detailed in Sec.~\ref{sec:membrane confidence}:

\begin{itemize}[leftmargin=*]
    \item \textit{Regression prediction}: it employs the timesteps and corresponding MAE pairs to conduct a regression prediction that models the MAE graph, which is subsequently used for estimating confidence in the scheduler for the upcoming $(t$+$1$)-th frame (⑦ in Fig.~\ref{fig:SNN_system_design}). 
    \item \textit{Confidence variance}: it calculates the cosine similarity between the spike features of the two most recent frames that did not utilize MPR, aiding in determining confidence variance, as discussed in Sec.~\ref{subsec:with} (⑧ in Fig.~\ref{fig:SNN_system_design}). 
\end{itemize}

\noindent
Finally, the calculated MAE for the MOD task is used to update the task's membrane confidence level (⑨ in Fig.~\ref{fig:SNN_system_design}).

\section{membrane confidence}
\label{sec:membrane confidence}

\hb{Based on multiple execution options (i.e., different timesteps w/o or w/ MPR) provided by the previous section to address \textsf{C1},
this section addresses \textsf{C2} by proposing a novel notion of \textit{membrane confidence} to capture the expected accuracy of timesteps w/o MPR (Sec.~\ref{subsec:without}) or w/ MPR (Sec.~\ref{subsec:with})}, which will be used for our scheduling framework to find the feasible execution combination offering the maximum accuracy \hb{without compromising timing guarantees} to address \textsf{C3} in Sec.~\ref{sec:scheduling}. 

To this end, we present the MAE metric, which measures the variation between sequential spike features over successive timesteps. 
For a MOD task $\tau_k$ with the given number of timesteps $\delta_k$ and timestep interval $g$ (e.g., 10)\footnote{The smaller values of $g$ may provide a more accurate approximation to the expected accuracy while it necessitates more frequent comparisons between spike features.}, MAE $M_k(\delta_k, t, g)$ is determined by aggregating the absolute disparities between each pair of corresponding elements in the spike features $s_k(\delta_k, t)$  and $s_k(\delta_k-g, t)$, and then averaging this sum over the total count of elements, which is formulated as

\noindent
\begin{equation}\label{eq:MAE}
    M_k(\delta_k, t, g) = |s_k(\delta_k, t) - s_k(\delta_k - g, t)|.    
\end{equation}

\noindent
Using MAE of $\tau_k$, we define the membrane confidence $\lambda_{k} (\delta_k, t, g)$ for the $t$-th frame of the MOD task $\tau_k$ for conducting the $\delta_k$ timesteps, determined as

\noindent
\begin{equation}\label{eq:membrane_confidence}
    \lambda_{k} (\delta_k, t, g) = 1 -max\Big(\frac{M_{k}(\delta_k, t, g) - M_{th}}{M_{k}(\delta^{min}-g, t, g) - M_{th}}, 0\Big),     
\end{equation}

\noindent
where $\delta^{min}$ (e.g., 50) is the minimum number of timesteps guaranteed for each task by offline schedulability analysis (to be detailed in Sec.~\ref{sec:scheduling}) without compromising timing guarantees, and $M_{th}$ (e.g., 0.02) denotes the predefined MAE threshold value required to achieve maximum accuracy.
$\lambda_{k} (\delta_k, t, g)$, ranging from 0 to 1, indicates the proximity of MAE at the minimum timestep $\delta^{min}$ (expressed as $M_{k}(\delta^{min}-g, t, g)$) to the established threshold $M_{th}$.

As depicted in Fig.~\ref{fig:mae_motivation_01}(a), MAE facilitates estimating the expected accuracy with the given timesteps.
However, the scheduling framework in tackling \textsf{C3} aims to determine, before the MOD execution for the $t$-th frame, the best combination of execution options to maximize accuracy (i.e., the highest membrane confidence). 
Hence, it becomes essential to estimate the attainable membrane confidence, inferred from the anticipated MAE, before executing the MOD task for the $t$-th frame. 
\hb{Sec.~\ref{subsec:without} presents the membrane confidence estimation for timesteps w/o MPR, while Sec.~\ref{subsec:with} delves into the estimation for timesteps w/ MPR.}

\subsection{Without membrane potential reuse}
\label{subsec:without}

To estimate the membrane confidence for the $t$-th frame of task $\tau_k$ with the number of timesteps $\delta_k$, we leverage the record of timesteps and associated MAE measurements from the latest execution of the MOD at time $t$-$f$, where the membrane potential was not reused.
To this end, we define a regression function $RF_k(x, t)$ that accepts timesteps $\delta_k$ at $t$ as input and provides the corresponding estimated MAE value as output. 
Considering the resemblance of the MAE graph's form to a rational function, as can be observed in Fig.~\ref{fig:mae_motivation_01}(a), we formulate the regression function $RF_k(x, t)$ as

\noindent
\begin{equation} \label{eq:regression}
    RF_k(x, t) = \frac{\alpha}{x} + \beta,
\end{equation}

\noindent
where $\alpha$ and $\beta$ are optimized using the Levenberg Marquardt method \cite{YNF01}, a least-squares approach that minimizes the residuals between timesteps and corresponding MAE values by leveraging the record of timesteps and associated MAE measurements from the ($t$-$f$)-th frame.
Utilizing the regression function $RF_k(x, t)$, we estimate the membrane confidence for the $t$-th frame with timesteps $\delta_k$ w/o MPR as

\noindent
\begin{equation}\label{eq:estimate_without_reuse}
    \overline{\lambda}_{k} (\delta_k, t, g) = 1 - max\Big(\frac{RF_{k}(\delta_k, t) - M_{th}}{RF_{k}(\delta^{min}-g, t) - M_{th}}, 0\Big).    
\end{equation}
\noindent

\begin{figure}
    \centering
    \includegraphics[width=1\linewidth]{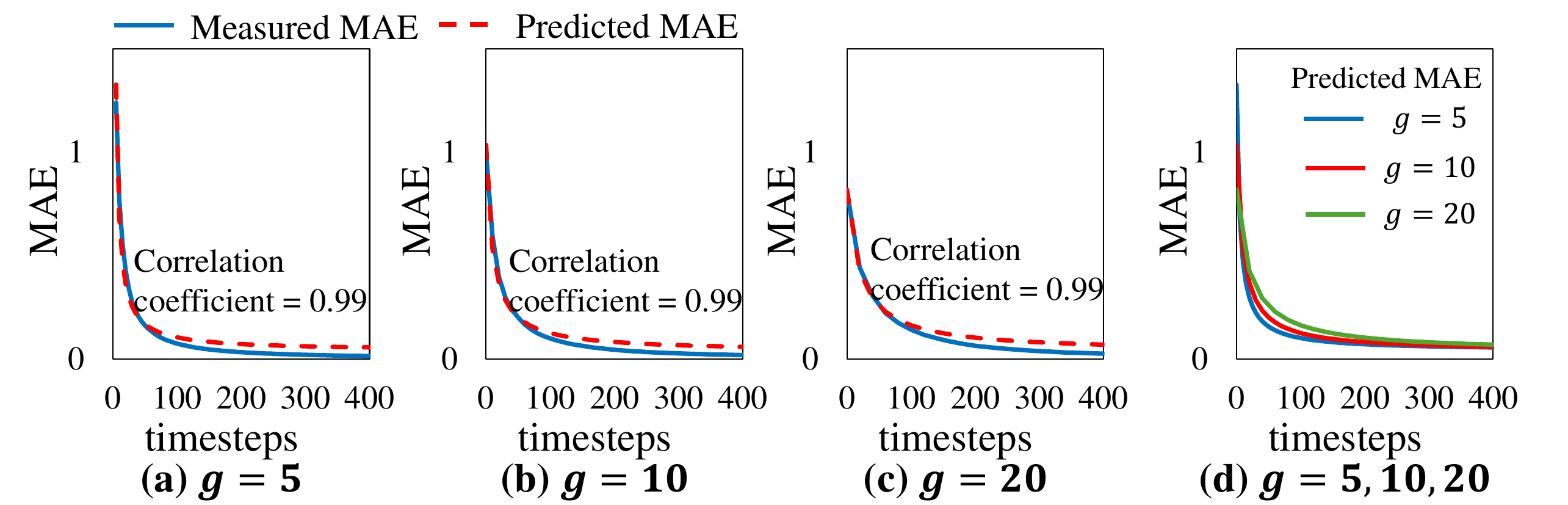}
    \caption{Precision on regression function $RF$ according to $g$}
    \label{fig:interval_exp}
    \vspace{-0.5cm}
\end{figure}

Fig.~\ref{fig:interval_exp} presents a strong resemblance between the actual MAE graph and our regression function $RF_k(x, t)$.
Figs.~\ref{fig:interval_exp}(a)--(c) display both the measured and the predicted MAE by the $RF$ function across 400 timesteps for various values of $g$. 
The $RF$ function is derived by initially measuring the MAE for 100 timesteps and subsequently predicting the MAE for an extended period of 400 timesteps. Our experimental data indicate a high correlation coefficient close to 0.99 for the measured and predicted MAE across all tested values of $g$ (i.e., $g=5, 10, 20$).
However, as illustrated in Figure~\ref{fig:interval_exp}(d), the accuracy of the $RF$ function varies with the parameter $g$. 
Specifically, a smaller $g$ results in more precise MAE predictions. 
This precision is particularly beneficial in environments where spike features change rapidly with the timesteps. 
Nevertheless, reducing $g$ also leads to more frequent comparisons between spike features, thereby potentially increasing the runtime overhead.

\subsection{With membrane confidence reuse}
\label{subsec:with}

In contrast to timesteps w/o MPR addressed in Sec.~\ref{subsec:without}, timesteps w/ MPR involve reusing the membrane potentials from the most recent execution in the ($t$-$f$)-th frame, where no MPR occurred. 
As discussed in Sec.~\ref{sec:intro}, it is essential to address (i) the expected accuracy (determined by MAE) of membrane potential extracted at $t$-$f$, (ii) its variance (determined by dynamics of the environment) introduced over $f$ frames, and (iii) the combined effect of (i), (ii), and the additional membrane potential extracted over given timesteps at $t$.

\noindent
For \hb{(i)}, Eq.~\eqref{eq:membrane_confidence} is utilized, by inputting $\delta_k$ for the ($t$-$f$)-th frame to compute the outcome straightforwardly. 
In addressing \hb{(ii)}, our focus is on the \textit{cosine distance}~\cite{BSY14} between spike features extracted at the most recent MOD execution without potential reuse at $t$-$f$ (denoted by $s(\delta_k^*, t$-$f)$ with timesteps $\delta_k^{t-f}$), and a previous one at $t$-$h$ (denoted by $s(\delta_k^*, t$-$h)$ with timesteps $\delta_k^{t-h}$) for $f < h$. 
Such the cosine distance enables the estimation of possible variance in membrane confidence for a single frame by dividing it by $h$-$f$. 
We then define $\delta_k^*$ as the smaller value between $\delta_k^{t-f}$ and $\delta_k^{t-h}$. 
Consequently, membrane confidence variance $\Delta \lambda_k(\delta_k^*, h, f)$ per each frame is approximated by
\resizebox{0.9\columnwidth}{!}{
\begin{minipage}{\columnwidth}
\begin{gather}   
    \Delta \lambda_k(\delta_k^*, h, f) = \frac{\gamma}{h-f} \cdot \Big(1 - max \Big(\frac{s(\delta_k^*, t-h)\cdot s(\delta_k^*, t-f)}{|s(\delta_k^*, t-h)||s(\delta_k^*, t-f)|}, 0\Big)\Big),
    \label{eq:diff}
\end{gather}
\end{minipage}
}

\noindent where $\gamma$ ($\ge 0$), 
set before the system begins operation, modulates the sensitivity of membrane confidence to the dynamics of spike features — the \kdh{lower} the $\gamma$, the higher the sensitivity. 
In scenarios where the ($t$-$h$)-th frame is absent, we default to computing $\Delta \lambda_k(\delta_k^*, h, f)$ using the inverse of $f$, which is $1/f$.

Then, to estimate the membrane confidence for the current $t$-th frame for \hb{(iii)}, we calculate a weighted sum of the membrane confidence $\lambda_{k} (\delta_k^{t-f}, t$-$f, g)$ from the reused ($t$-$f$)-th frame, accounting for its variance $\Delta \lambda_k(\delta_k^*, h, f)$ over $f$ frames, and the confidence gain achievable from $\delta_{k}$ additional timesteps in the current $t$-th frame, which is formulated as

\resizebox{\columnwidth}{!}{
\begin{minipage}{\columnwidth}
\begin{align}
    &\overline{\lambda}^{+}_{k}  (\delta_k, t, g) = \lambda_{k} (\delta_k^{t-f}, t-f, g) \cdot max\big(1 -  \Delta \lambda_k(\delta_k^*, h, f) \times f, 0\big) \nonumber\\ 
    &+ \big(1 - \lambda_{k} (\delta_k^{t-f}, t-f, g)\big) \cdot \bigg(1 - max\bigg(\frac{RF_{k}(\delta_k^{t-f} + \delta_k, t) - M_{th}}{RF_{k}(\delta_k^{t-f}, t) - M_{th}}, 0\bigg)\bigg).
    \label{eq:estimate_with_reuse}
\end{align}
\end{minipage}
}

\noindent
Using the membrane potential from the ($t$-$f$)-th frame implies an assumption that $\delta_k^{t-f}$ timesteps have already been performed (without computation at $t$). Thus, Eq.~\eqref{eq:estimate_with_reuse} employs Eq.~\eqref{eq:regression} to estimate the extent to which $RF_{k}(\delta_k^{t-f}+\delta_k, t)$ approaches $M_{th}$ relative to $RF_{k}(\delta_k^{t-f}, t)$, for determining the confidence gain achievable from $\delta_{k}$ additional timesteps.

\section{Scheduling Framework for \sys{}}
\label{sec:scheduling}

We consider an SNN task set $\tau$, which consists of $n$ SNN tasks $\tau_i \in \tau$.
Each task $\tau_{i}$ is represented by $\tau_{i}=(T_{i}, C_{i}(\delta))$,
where $T_i$ denotes the period of $\tau_i$ and $C_i(\delta)$ denotes the worst-case execution time of $\tau_i$ when the number of timesteps to be executed is $\delta$.
The latter is 
calculated by $C_i(\delta) = \delta \times c_i^{ts} + c_i^{fc}$, where $c_i^{ts}$ represents the worst-case execution time for performing a single timestep, and $c_i^{fc}$ is the time taken to execute a final convolutional layer;
therefore, $C_i(\delta)$ increases linearly with $\delta$, the number of timesteps to be executed.
Each job $J_i$, invoked by task $\tau_i$, is released every $T_i$ and must complete its execution within $T_i$, which is a timing constraint. We use task $\tau_i$ and job $J_i$ interchangeably when no ambiguity arises.

As a base scheduling algorithm,
\sys{} employs \npfp{} (Non-Preemptive Fixed-Priority scheduling with MINimum execution)~\cite{KLC22} as follows.
By ``NP'', any job, once starting its execution, is not preempted by any other job until its execution is completed.
By ``FP'', each task $\tau_i$ has a pre-determined task-level priority, which is inherited to every job invoked by $\tau_i$; let LP($\tau_{i}$) and HP($\tau_{i}$) denote a set of tasks with lower and higher priorities than $\tau_{i}$, respectively.
By ``MIN'', every job of $\tau_i$ executes for the minimum duration denoted by $C_i(\delta^{min})$,
where $\delta^{min}$ is the timesteps required to achieve the minimum expected accuracy (determined by the system designer) provided by the system.
The following lemma derives an offline schedulability analysis for \npfp{}.

\lemma{\label{lemma:npfpmin}Suppose that an SNN task set $\tau$ is scheduled by \npfp{}. If every $\tau_{i} \in \tau$ satisfies Eq.~\eqref{eq:offline_analysis}, the following holds for every $\tau_{i} \in \tau$: any job invoked by $\tau_{i}$ does not miss its deadline.}

\noindent
\begin{align}
     &C_{i}(\delta^{min})+\max_{\tau_j\in \textsf{LP}(\tau_i)} C_{j}(\delta^{min}) \nonumber\\&
     +\sum_{\tau_h\in\textsf{HP}(\tau_i)} \bigg\lceil\frac{T_i+T_h-C_{h}(\delta^{min})}{T_h}\bigg\rceil\cdot C_{h}(\delta^{min}) \le T_i.
     \label{eq:offline_analysis}
\end{align}

\begin{IEEEproof}
The lemma comes from \cite{BCL09,GRS96}. A lower-priority job can block a job of $\tau_i$ of interest (denoted by $J_i$), only if the lower-priority job starts its execution before $J_i$ is released; this is addressed in $\max_{\tau_j\in \textsf{LP}(\tau_i)} C_{j}(\delta^{min})$~\cite{GRS96}.
Also, 
the amount of execution of jobs of a higher-priority task $\tau_h$ in an interval of length $T_i$ cannot be larger than $\big\lceil\frac{T_i+T_h-C_{h}(\delta^{min})}{T_h}\big\rceil\cdot C_{h}(\delta^{min})$~\cite{BCL09}. Therefore, if the LHS (Left-Hand-Side) of Eq.~\eqref{eq:offline_analysis} is not larger than $T_i$, the job of $\tau_i$ of interest can finish its execution within the interval of length $T_i$.
\end{IEEEproof}

\begin{algorithm} [t]
\caption{The \npfpmem{} scheduling algorithm}
\label{algo:online_scheduling_algorithm}
\small
\raggedright
At $t$, at which the highest-priority active job of $\tau_k$ starts:

\begin{algorithmic}[1]
    \STATE Let $\chi$ be the set of the active tasks (except $\tau_k$) at $t$ and the inactive tasks each of whose 
    nearest future job will be released in $[t, r_{k}^{(t)})$. 
    \STATE $\Delta \delta = \big(r_{k}^{(t)} - t - C_{k}(\delta^{min})\big)/c_i^{ts}$
    \FOR{$\tau_i \in \chi$}
        \STATE $\Delta \delta = min(\Delta \delta_{i}^{(t)}, \Delta \delta$)
    \ENDFOR   
    \IF{$\overline{\lambda}_{k}(\delta^{min} + \Delta \delta, t, g) \ge \overline{\lambda}^{+}_{k}(\delta^{min} + \Delta \delta, t, g)$} 
        \STATE Execute $\tau_{k}$ for $C_{k}(\delta^{min} + \Delta \delta)$ 
    \ELSE
        \STATE Execute $\tau_{k}$ for $C_{k}(\delta^{min} + \Delta \delta)$ with MPR 
    \ENDIF    
    \FOR{$\tau_i \in \chi$}
        \STATE $\Delta \delta_{i}^{(t)} = \Delta \delta_{i}^{(t)} - \Delta \delta$
    \ENDFOR
    \STATE $\Delta \delta_k^{(t)}$ is set to \jl{$T_i-$ the LHS of Eq.~\eqref{eq:offline_analysis}}
\end{algorithmic}
\normalsize
\end{algorithm}

Although \npfp{} can provide timing guarantees subject to Lemma~\ref{lemma:npfpmin},
it cannot maximize the accuracy of SNN tasks as it disallows each job of $\tau_i$ to be executed for more than $C_i(\delta^{min})$.
On top of \npfp{}, we propose \npfpmem{} 
\jl{(Non-Preemptive Fixed-Priority Scheduling with MEMbrane confidence)}
in Algo.~\ref{algo:online_scheduling_algorithm}, which establishes two principles:
(\textsf{P1}) no deadline miss guaranteed if Lemma~\ref{lemma:npfpmin} holds,
and 
(\textsf{P2}) accuracy maximization achieved by the result of Sec.~\ref{sec:membrane confidence}.

At $t$, at which the highest-priority active job of $\tau_k$ starts,\footnote{\jl{A job is said to be \textit{active} at $t$, if the job has remaining execution at $t$.}} Algo.~\ref{algo:online_scheduling_algorithm} focuses on the interval of $[t,r_k^{(t)})$, where $r_k^{(t)}$ denotes the earliest deadline of any job of $\tau_k$ after $t$ (therefore, $r_k^{(t)}$ also implies the next job release time of $\tau_k$).
Then, it calculates the number of timesteps $\Delta\delta$ additionally executed without compromising the schedulability of tasks affected by the additional execution.
To this end, every task $\tau_i$ maintains $\Delta\delta_i^{(t)}$, indicating how many timesteps are allowed to be additionally executed for its individual job active at $t$.  

The process begins by determining the set of tasks, denoted as $\chi$, whose schedulability is influenced by the additional execution (Line 1). Next, it calculates the number of additional timesteps available for the task $\tau_k$ of interest, initially setting this value to $\Delta\delta$ (Line 2). Following this, $\Delta\delta$ is updated by taking the minimum value among $\Delta\delta_i^{(t)}$ for every task $\tau_i \in \chi$ (Lines 3--5). Then, the expected accuracy with and without MPR is calculated, and the execution option with higher accuracy is selected (Lines 6--10). After that, $\Delta\delta_i^{(t)}$ is updated for every task $\tau_i \in \chi$ by considering the additional execution of $\Delta\delta$ timesteps (Lines 11--13). Finally, $\tau_k$ resets $\Delta\delta_k^{(t)}$ for its next job to the difference between the LHS and RHS (Right-hand-side) of Eq.~\eqref{eq:offline_analysis}, as any job of $\tau_k$ does not miss its deadline even with additional execution for the difference according to Eq.~\eqref{eq:offline_analysis} (Line 14). Note that \sys{} uses $RF_{k}(\Delta\delta, t)$ to approximately upper-bound $\Delta\delta$ not to exceed the corresponding timesteps of $M_{th}$ to avoid execution beyond the timestep for reaching $M_{th}$.

\npfpmem{} in Algo.~\ref{algo:online_scheduling_algorithm} achieves \textsf{P1} and \textsf{P2} as follows.

\theorem{\label{theorem:npfpmem}
Suppose that an SNN task set $\tau$ is scheduled by \npfpmem{}. If every $\tau_{i} \in \tau$ satisfies Eq.~\eqref{eq:offline_analysis}, the following two properties hold:
(offline timing guarantees) for every $\tau_i\in\tau$, 
any job invoked by $\tau_{i}$ does not miss its deadline,
and (run-time accuracy improvement) the accuracy of every job under \npfpmem{} is equal to or higher than that under \npfp{}.}
\begin{IEEEproof}
Since additional execution beyond $C_i(\delta^{min})$ is performed only if it does not compromise the schedulability of any other job (by Lines 2--5), the offline timing guarantees in Lemma~\ref{lemma:npfpmin} still holds for \npfpmem{}.
Since the number of timesteps to be executed for $\tau_k$ is $\delta^{min}+\Delta\delta \ge \delta^{min}$ (by Lines~7 and 9), 
the run-time accuracy improvement is achieved.
\end{IEEEproof}

\hb{
\textbf{Run-time complexity of \npfpmem{}.}
At each scheduling decision at $t$ (when a job is either completed or released), \npfpmem{} must calculate $\Delta \delta$. 
Let $|\chi|$ denote the number of tasks in the set $\chi$ as shown in Algo.~\ref{algo:online_scheduling_algorithm}; note that $|\chi|$ is equal to or less than the total number of tasks. 
\npfpmem{} processes Line 4 and Line 12 of Algo.~\ref{algo:online_scheduling_algorithm} exactly $|\chi|$ times each. 
Therefore, the complexity of \npfpmem{} is $O(|\chi|)$ at each scheduling decision at $t$.
}

\section{Evaluation}
\label{sec:evaluation}

\textbf{Hardware and software.} 
All experiments are conducted on a server featuring AMD EPYC 7302 16-Core processors, 198GB RAM, and an NVIDIA RTX A6000 48GB GPU, operating on Ubuntu 18.04.6 with CUDA 11.4. 
These experiments employ an SNN simulator based on Python 3.6.9, aligned with the same simulation environment of Spiking-YOLO~\cite{KPN20}
\hb{as Spiking-YOLO on a neuromorphic chip of TrueNorth~\cite{MAA14} demonstrates 280 times more energy efficiency than Tiny-YOLO on a GPU of NVIDIA
Tesla V100.}
The detection process involved Spiking-YOLO trained for 300 epochs on the MS COCO 2014 dataset~\cite{LMB14}.
Our experiments concentrate exclusively on the \jll{single-timestep} execution time of $c_{i}^{ts}$ for each $\tau_{i} \in \tau$, given that the computational cost $c_{i}^{fc}$ \jll{(for the final convolutional layer)} is relatively insignificant.\footnote{With a timestep of 50 in our experiment, the number of operations for the final convolutional layer amounts to a mere 0.04\% of the total number of operations.}
We conduct a performance evaluation using the KITTI dataset~\cite{APR12} and utilize the mAP metric~\cite{JRM18} as a standard for comparing the detection efficacy of different approaches.
\hb{As \sys{} accepts any fixed-priority scheduling, we deploy rate-monotonic (RM)~\cite{LiLa73} for our experiment.}

\textbf{Considered approaches.}
To evaluate performance, we compare the following approaches.

\begin{itemize} [leftmargin=*]
    \item \textbf{$\texttt{MIN}$}: \npfp{} \hb{(i.e., w/o MPR)} with the largest $\delta^{min}$ that satisfies Eq.~\eqref{eq:offline_analysis}.\footnote{Note that this operates identically to the existing Spiking-YOLO \jl{subject to} 
    the timing constraint corresponding to Eq.~\eqref{eq:offline_analysis}.}
    \item \textbf{$\texttt{MIN+}$}: $\texttt{MIN}$ 
    w/ MPR for every two frames.
    \item \textbf{$\texttt{MEM}$}: \npfpmem{} (in Algo.~\ref{algo:online_scheduling_algorithm}, \hb{i.e., w/ MPR}) with $\delta^{min}$=50 (our ultimate approach).
    \item \textbf{$\texttt{MEM-}$}: $\texttt{MEM}$ w/o MPR.
\end{itemize}

\noindent
For a favorable comparison with $\texttt{MIN}$ and $\texttt{MIN+}$, we assigned the largest $\delta^{min}$ 
\jl{(subject to 
Eq.~\eqref{eq:offline_analysis})} to $\texttt{MIN}$ and $\texttt{MIN+}$, as a $\delta^{min}$ of 50 yielded much lower accuracy.

\begin{figure}[t]
    \centering
    \subfloat[$\delta^{min}$=$80$]{\includegraphics[width=0.5\linewidth]{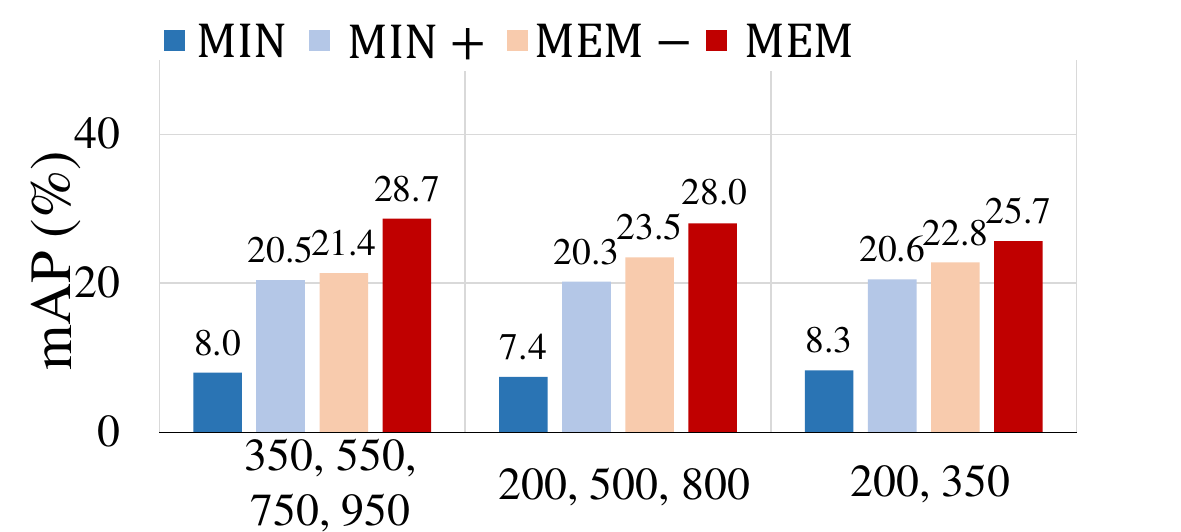}}
    \subfloat[$\delta^{min}$=$120$]{\includegraphics[width=0.5\linewidth]{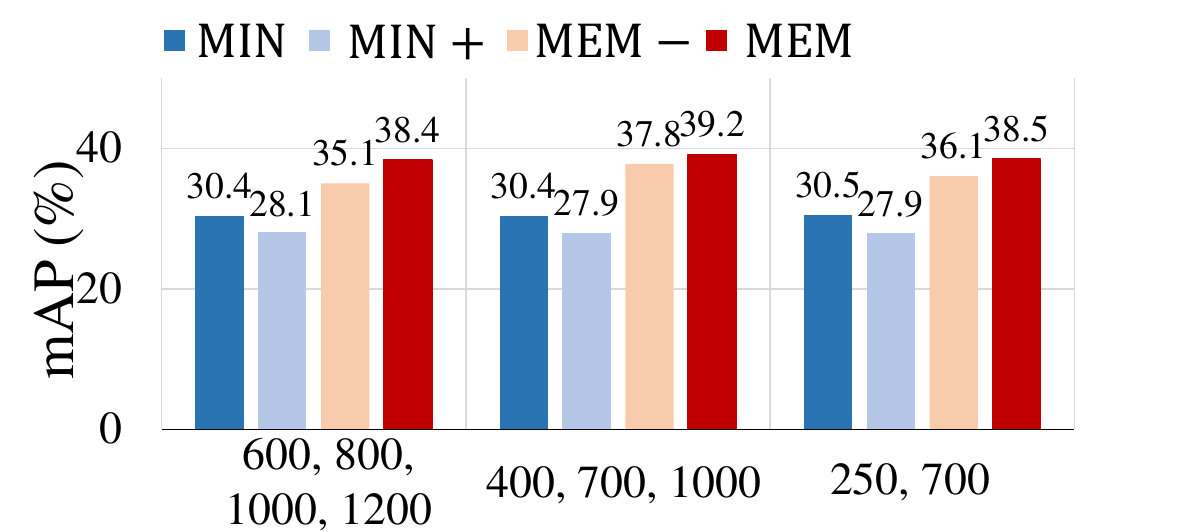}}
    \caption{Experiment results on various approaches}
    \label{fig:evaluation_kitti}
    \vspace{-0.6cm}
\end{figure}

\textbf{General accuracy.} We first evaluate the accuracy of the considered approaches for $n$ ranging from 2 to 4 with different periods (denoted by the x-axis in each figure), setting $g$=$10$, $\gamma$=$3$ and \kdh{$M_{th}$=$0.08$} as experiment parameters.
Fig.~\ref{fig:evaluation_kitti}(a) presents the accuracy for both $\texttt{MIN}$ and $\texttt{MIN+}$ at $\delta^{\min}$=$80$ (and for $\texttt{MEM-}$ and $\texttt{MEM}$ at $\delta^{\min}$=$50$ as a default value). 
Notably, $\texttt{MIN}$ records the least accuracy due to limited timesteps. 
However, $\texttt{MIN+}$ highlights the benefit of alternating MPR in enhancing accuracy.
Furthermore, $\texttt{MEM-}$ surpasses both $\texttt{MIN}$ and $\texttt{MIN+}$ for all task sets in accuracy by leveraging $\Delta \delta$ during runtime. 
Finally, $\texttt{MEM}$ illustrates that employing MPR can yield even greater accuracy. 
Fig.~\ref{fig:evaluation_kitti}(b) reveals that with $\delta^{\min}$=$120$, $\texttt{MIN+}$ shows reduced accuracy compared to $\texttt{MIN}$, suggesting that MPR without membrane confidence awareness can lower accuracy at higher timesteps.
Conversely, $\texttt{MEM}$ attains superior accuracy by dynamically reusing membrane potentials as needed, guided by membrane confidence.

\begin{figure}[t]
    \centering
    \subfloat[Various $M_{th}$]{\includegraphics[width=0.5\linewidth]{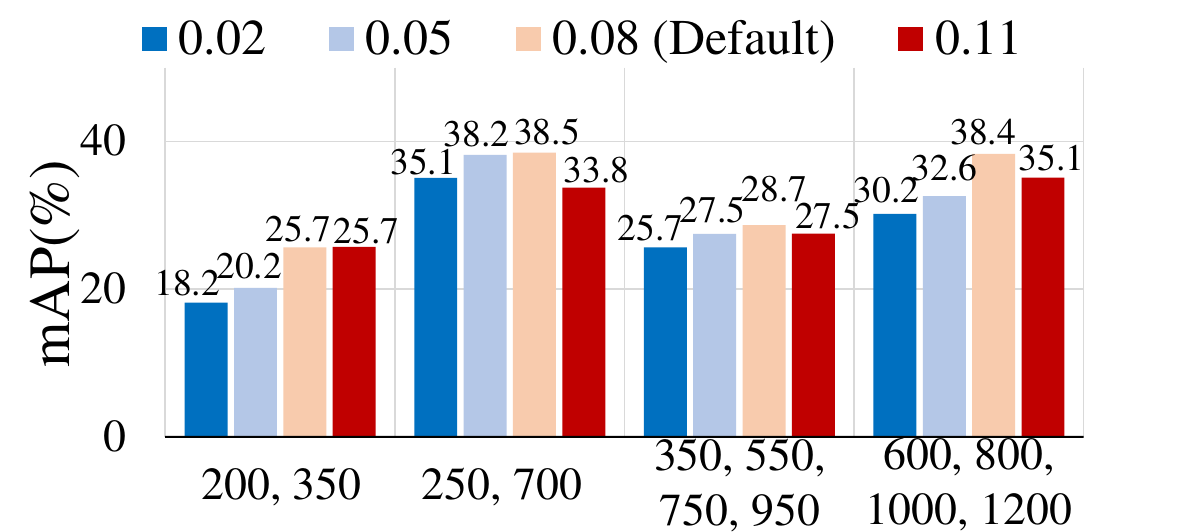}}
    \subfloat[Various $\gamma$]{\includegraphics[width=0.5\linewidth]{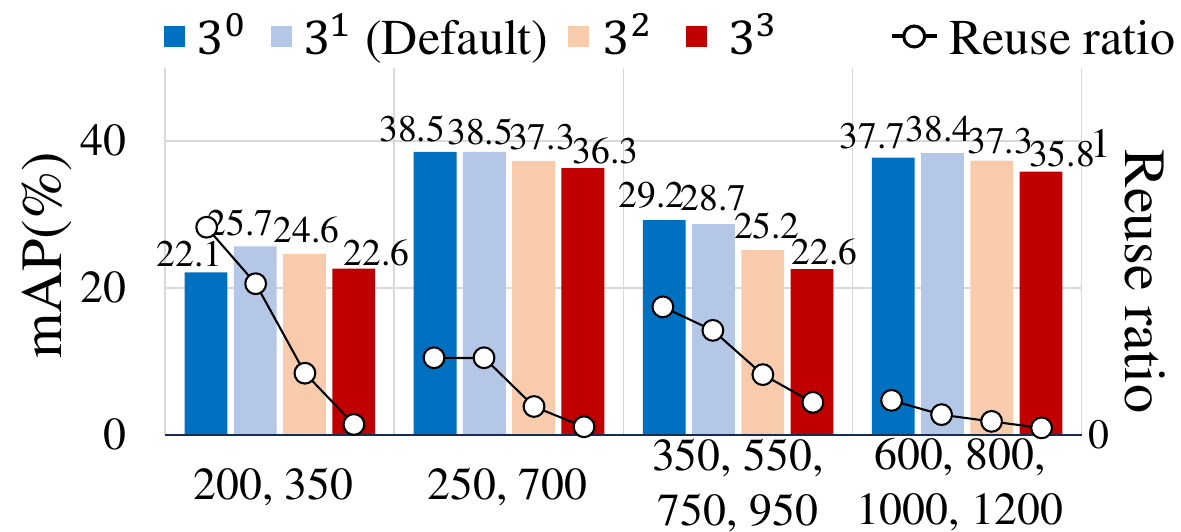}}
    \caption{Experiment results on \sys{} with various parameters}
    \label{fig:ablation_kitti}
    \vspace{-0.3cm}
\end{figure}

\remove{
\begin{figure}[t]
    \centering
    \subfloat[Various $M_{th}$ when $n$=2]{\includegraphics[width=0.5\linewidth]{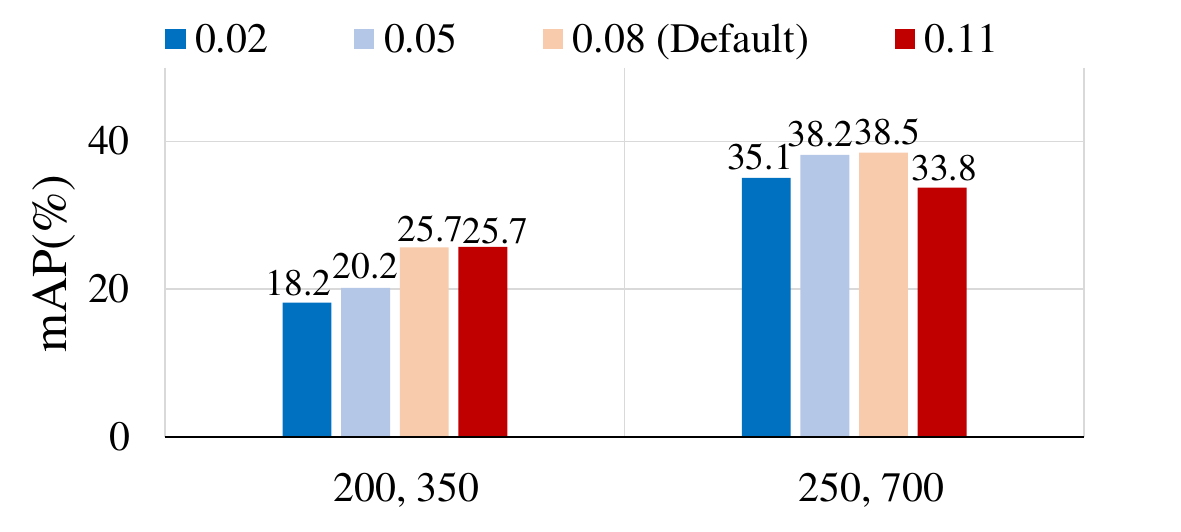}}
    \subfloat[Various $M_{th}$ when $n$=4]{\includegraphics[width=0.5\linewidth]{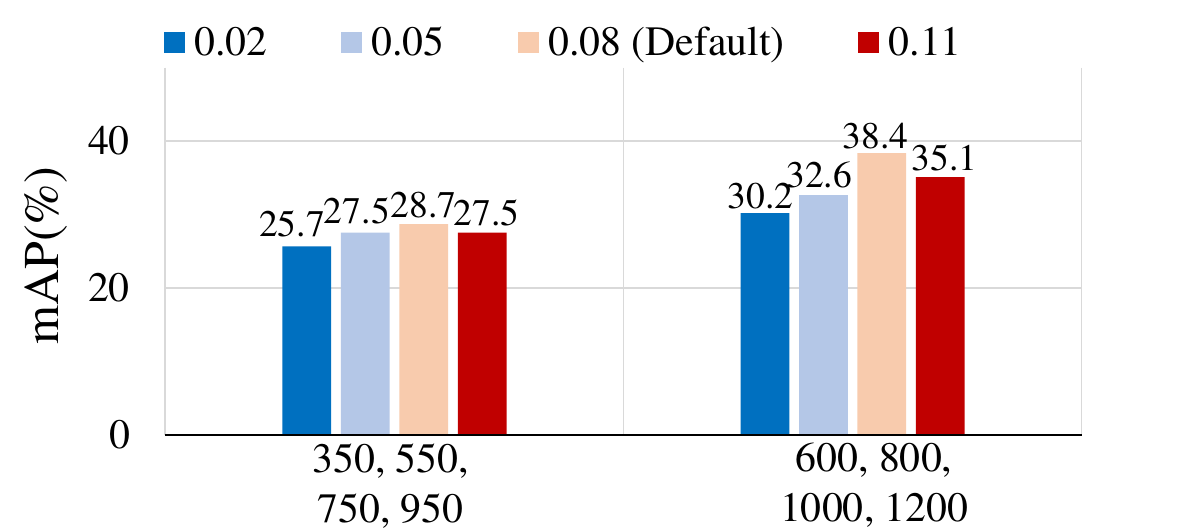}} \\
    \subfloat[Various $\gamma$ when $n$=2]{\includegraphics[width=0.5\linewidth]{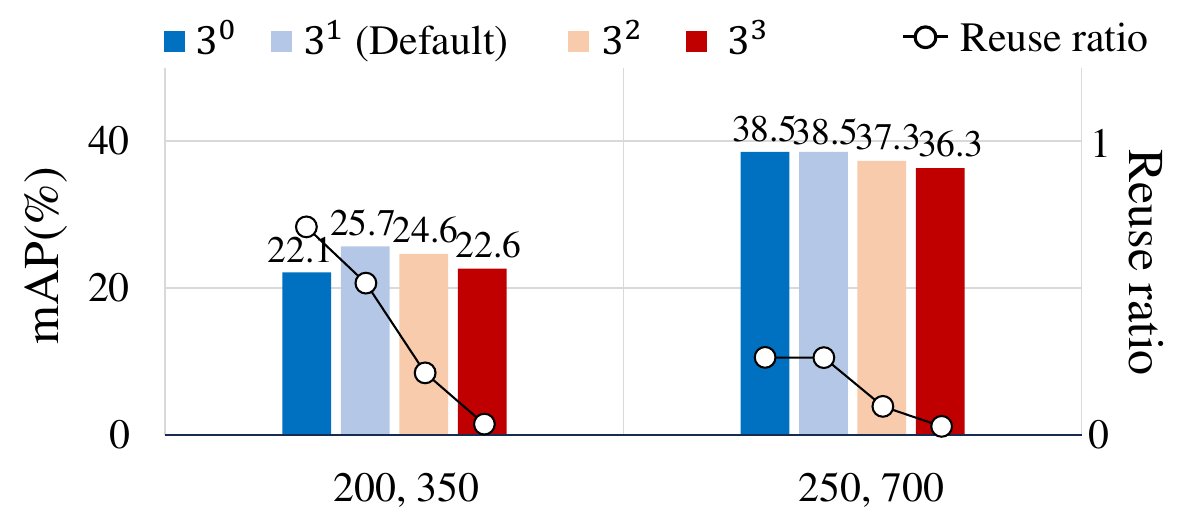}}
    \subfloat[Various $\gamma$ when $n$=4]{\includegraphics[width=0.5\linewidth]{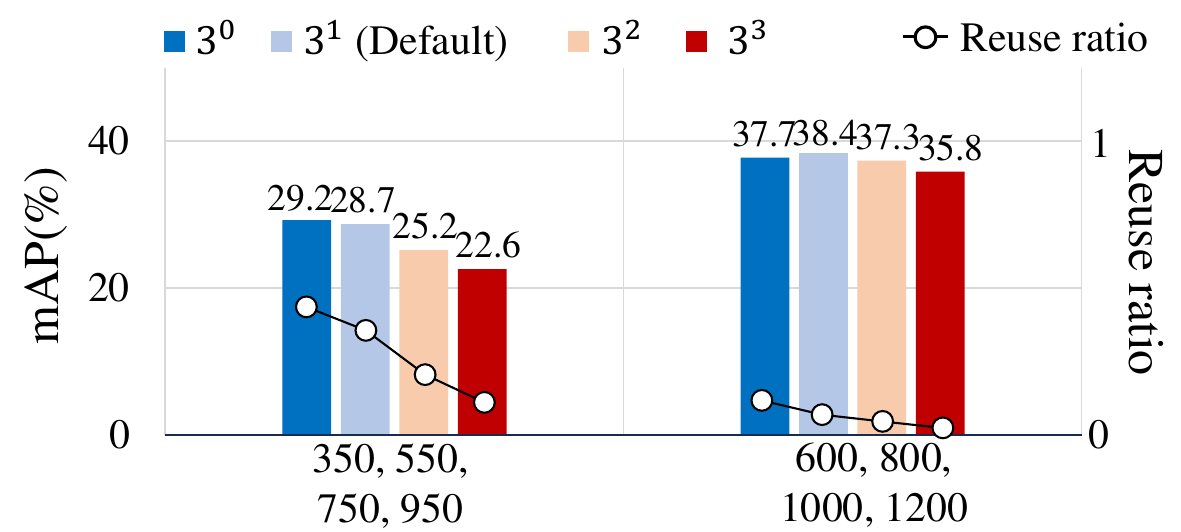}} \\
    \caption{Experiment results on \sys{} with various parameters
    }
    \label{fig:ablation_kitti}
    \vspace{-0.3cm}
\end{figure}
}

\textbf{Impact of various $M_{th}$.}
Each SNN execution on \sys{} performs until the timestep reaches $M_{th}$, enabling the calibration of computational resources for each task. 
Fig.~\ref{fig:ablation_kitti}(a) illustrates the varying maximum, minimum, and average accuracy depending on the number of tasks. 
\hb{A threshold $M_{th}$ set too low (e.g., 0.02) hinders equitable resource distribution across multiple tasks by enabling the allocation of a disproportionately large share of resources to a single task during each scheduling decision, potentially diminishing the overall accuracy. 
Conversely, setting the threshold $M_{th}$ excessively high (e.g., 0.11) can also reduce the overall accuracy due to the resultant underutilization of resources across multiple tasks.
Fig.~\ref{fig:ablation_kitti}(a) illustrates that the optimal accuracy across various $M_{th}$ values is achieved at an $M_{th}$ of 0.08.
}

\textbf{Impact of various $\gamma$.}
Membrane confidence estimator adjusts sensitivity to dynamics of the environment using $\gamma$ in Eq.~\eqref{eq:diff}.
Fig.~\ref{fig:ablation_kitti}(b) shows mAP and reuse \hb{(of membrane potential)} ratio depending on $\gamma$ when there are 2 and 4 tasks, respectively.
In all task sets, as $\gamma$ decreases, the sensitivity to the dynamics of the environment decreases, and the reuse ratio increases.
High-frequency \kdh{MPR} does not capture the dynamics of the environment due to accumulated membrane potentials from previous frames, while low-frequency \kdh{MPR} does not ensure sufficient timesteps to detect objects.
For example, with the task set $(600 ,800 ,1000 ,1200)$, \sys{} achieves its highest mAP at $3^{1}$.
However, deviating from $3^{1}$--either by reusing more or less--leads to a decrease in mAP.

\begin{table}[t]
\caption{Energy efficiency comparison}
\resizebox{\columnwidth}{!}{%
\footnotesize
\begin{tabular}{c|c|cccc}
\hline
Period set                & Approach                  &Avg. timesteps     & mAP(\%)                   & Energy                         \\ \hline
\multirow{3}{*}{\footnotesize{170, 500}} & $\texttt{MEM}$            & 75.1               & \textbf{\underline{22.2}} & \textbf{\underline{1x}}        \\ \cline{2-6} 
                          & $\texttt{MIN}_\texttt{E}$ & 80                 & 7.2                       & \textbf{\underline{1.02x}}     \\ \cline{2-6} 
                          & $\texttt{MIN}_\texttt{A}$ & 103                & \textbf{\underline{21.7}} & 1.32x                          \\ \hline
\multirow{3}{*}{\footnotesize{200, 700}} & $\texttt{MEM}$            & 118.6              & \textbf{\underline{32.5}} & \textbf{\underline{1x}}        \\ \cline{2-6} 
                          & $\texttt{MIN}_\texttt{E}$ & 121                & 30.3                      & \textbf{\underline{1.01x}}     \\ \cline{2-6} 
                          & $\texttt{MIN}_\texttt{A}$ & 126                & \textbf{\underline{32.5}} & 1.05x                          \\ \hline
\multirow{3}{*}{\footnotesize{300, 400, 500, 600}} & $\texttt{MEM}$  & 67.4               & \textbf{\underline{16.5}} & \textbf{\underline{1x}}        \\ \cline{2-6} 
                          & $\texttt{MIN}_\texttt{E}$ & 70                 & 3.8                       & \textbf{\underline{1x}}        \\ \cline{2-6} 
                          & $\texttt{MIN}_\texttt{A}$ & 93                 & \textbf{\underline{15.7}} & 1.32x                          \\ \hline
\multirow{3}{*}{\footnotesize{400, 450, 500, 550}} & $\texttt{MEM}$  & 77.3               & \textbf{\underline{18.6}} & \textbf{\underline{1x}}        \\ \cline{2-6} 
                          & $\texttt{MIN}_\texttt{E}$ & 81                 & 8.8                       & \textbf{\underline{1.02x}}     \\ \cline{2-6} 
                          & $\texttt{MIN}_\texttt{A}$ & 95                 & \textbf{\underline{17.1}} & 1.19x                          \\ \hline
\end{tabular}%
}
\label{tab:energy_table}
\normalsize
\vspace{-0.4cm}
\end{table}

\textbf{Energy consumption.} 
Theoretical power consumption in SNN is defined by the following formula generally recognized in state-of-the-art SNN studies~\cite{KPN20}.

\noindent
\begin{equation}
    E = \delta \times f_{r} \times E_{AC} \times OP_{AC} + E_{MAC} \times OP_{MAC},
    \label{eq:energy_calc}
\end{equation}

\noindent
where $\delta$ is the number of repeated timesteps, $f_{r}$ is the firing ratio, and $OP_{MAC}$ and $OP_{AC}$ are the counts of multiply-and-accumulate and spike-based accumulate operations, respectively. 
The energy requirements per operation, $E_{AC}$ and $E_{MAC}$, are theoretically estimated at $0.9pJ$ for $OP_{AC}$ and $4.6pJ$ for $OP_{MAC}$~\cite{HOM14}.
To assess $\texttt{MEM}$'s power efficiency against conventional Spiking-YOLO (i.e., $\texttt{MIN}$), we measured mAP and energy consumption (Eq.~\eqref{eq:energy_calc}) for each task in the period sets of \kdh{(170, 500), (200, 700), (300, 400, 500, 600) and (400, 450, 500, 550)}.
Additionally, setting aside timing guarantees, we compare two approaches, $\texttt{MIN}_\texttt{E}$ and $\texttt{MIN}_\texttt{A}$, with $\texttt{MEM}$, emphasizing equivalence in energy usage and accuracy.
\kdh{For $\texttt{MIN}_\texttt{E}$ and $\texttt{MIN}_\texttt{A}$, $\delta^{min}$ was uniformly increased across datasets to match $\texttt{MEM}$'s energy and accuracy respectively.}
Table~\ref{tab:energy_table} displays each method's average timesteps, mAP, and energy use. 
It is shown from Table~\ref{tab:energy_table} that $\texttt{MEM}$, leveraging \hb{MPR} and membrane confidence, attains higher accuracy than $\texttt{MIN}_\texttt{E}$ and consumes less energy than $\texttt{MIN}_\texttt{A}$. 
Significantly, $\texttt{MEM}$ shows enhanced power efficiency and accuracy in shorter periods, attributed to its proficiency in attaining high accuracy with fewer timesteps through MPR.
\kdh{Also note that $\texttt{MIN}_\texttt{E}$ (except the period set $(170, 500)$) and $\texttt{MIN}_\texttt{A}$ do not guarantee timely execution for all tasks, whereas $\texttt{MEM}$ guarantee timely execution for all tasks listed in Table.~\ref{tab:energy_table}.}

\section{Conclusion}
\label{sec:conclusion}

In this paper, we proposed \sys{}, the first system ensuring \textsf{R1} and \textsf{R2} in SNN-based MOD systems on AMAs. 
\sys{} provides two execution options—timestep and frame—with flexible timesteps, w/o or w/ MPR, for multiple SNN-based MOD tasks to support \textsf{R1}. 
Additionally, \sys{} incorporates a predictive measure for detection accuracy variation based on execution options, supporting \textsf{R2}. Furthermore, \sys{} introduces a scheduling framework with offline schedulability analysis and a runtime algorithm for maximizing accuracy via membrane confidence. We implemented RT-SNN with Spiking-YOLO, an SNN-based MOD model, and our experiments demonstrate \sys{}'s effectiveness in meeting both \textsf{R1} and \textsf{R2} while maintaining SNN energy efficiency.

\bibliographystyle{IEEEtran}
\bibliography{main}

\begin{thebibliography}{10}
\providecommand{\url}[1]{#1}
\csname url@samestyle\endcsname
\providecommand{\newblock}{\relax}
\providecommand{\bibinfo}[2]{#2}
\providecommand{\BIBentrySTDinterwordspacing}{\spaceskip=0pt\relax}
\providecommand{\BIBentryALTinterwordstretchfactor}{4}
\providecommand{\BIBentryALTinterwordspacing}{\spaceskip=\fontdimen2\font plus
\BIBentryALTinterwordstretchfactor\fontdimen3\font minus \fontdimen4\font\relax}
\providecommand{\BIBforeignlanguage}[2]{{%
\expandafter\ifx\csname l@#1\endcsname\relax
\typeout{** WARNING: IEEEtran.bst: No hyphenation pattern has been}%
\typeout{** loaded for the language `#1'. Using the pattern for}%
\typeout{** the default language instead.}%
\else
\language=\csname l@#1\endcsname
\fi
#2}}
\providecommand{\BIBdecl}{\relax}
\BIBdecl

\bibitem{KCK22}
W.~Kang, S.~Chung, J.~Y. Kim, Y.~Lee, K.~Lee, J.~Lee, K.~G. Shin, and H.~S. Chwa, ``{DNN-SAM}: Split-and-merge dnn execution for real-time object detection,'' in \emph{IEEE Real Time Technology and Applications Symposium (RTAS)}, 2022, pp. 160--172.

\bibitem{KLC22}
D.~Kang, S.~Lee, H.~S. Chwa, S.-H. Bae, C.~M. Kang, J.~Lee, and H.~Baek, ``Rt-mot: Confidence-aware real-time scheduling framework for multi-object tracking tasks,'' in \emph{IEEE Real-Time Systems Symposium (RTSS)}.\hskip 1em plus 0.5em minus 0.4em\relax IEEE, 2022, pp. 318--330.

\bibitem{DSF21}
S.~Davidson and S.~B. Furber, ``Comparison of artificial and spiking neural networks on digital hardware,'' \emph{Frontiers in Neuroscience}, vol.~15, p. 651141, 2021.

\bibitem{PRVS23}
R.~V.~W. Putra and M.~Shafique, ``Topspark: a timestep optimization methodology for energy-efficient spiking neural networks on autonomous mobile agents,'' \emph{arXiv preprint arXiv:2303.01826}, 2023.

\bibitem{PVS23}
------, ``Mantis: enabling energy-efficient autonomous mobile agents with spiking neural networks,'' in \emph{2023 9th International Conference on Automation, Robotics and Applications (ICARA)}.\hskip 1em plus 0.5em minus 0.4em\relax IEEE, 2023, pp. 197--201.

\bibitem{MAA14}
P.~A. Merolla, J.~V. Arthur, R.~Alvarez-Icaza, A.~S. Cassidy, J.~Sawada, F.~Akopyan, B.~L. Jackson, N.~Imam, C.~Guo, Y.~Nakamura \emph{et~al.}, ``A million spiking-neuron integrated circuit with a scalable communication network and interface,'' \emph{Science}, vol. 345, no. 6197, pp. 668--673, 2014.

\bibitem{KPN20}
S.~Kim, S.~Park, B.~Na, and S.~Yoon, ``Spiking-yolo: spiking neural network for energy-efficient object detection,'' in \emph{Proceedings of the AAAI conference on artificial intelligence (AAAI)}, vol.~34, no.~07, 2020, pp. 11\,270--11\,277.

\bibitem{WSZ21}
T.~Wang, C.~Shi, X.~Zhou, Y.~Lin, J.~He, P.~Gan, P.~Li, Y.~Wang, L.~Liu, N.~Wu \emph{et~al.}, ``Compsnn: A lightweight spiking neural network based on spatiotemporally compressive spike features,'' \emph{Neurocomputing}, vol. 425, pp. 96--106, 2021.

\bibitem{KPNKY20}
S.~Kim, S.~Park, B.~Na, J.~Kim, and S.~Yoon, ``Towards fast and accurate object detection in bio-inspired spiking neural networks through bayesian optimization,'' \emph{IEEE Access}, vol.~9, pp. 2633--2643, 2020.

\bibitem{SCH23}
Q.~Su, Y.~Chou, Y.~Hu, J.~Li, S.~Mei, Z.~Zhang, and G.~Li, ``Deep directly-trained spiking neural networks for object detection,'' in \emph{In Proceedings of the IEEE/CVF International Conference on Computer Vision (ICCV)}, 2023, pp. 6555--6565.

\bibitem{APR12}
A.~Geiger, P.~Lenz, and R.~Urtasun, ``Are we ready for autonomous driving? the kitti vision benchmark suite,'' in \emph{In Proceedings of the IEEE/CVF Conference on Computer Vision and Pattern Recognition (CVPR)}, 2012.

\bibitem{YNF01}
N.~Yamashita and M.~Fukushima, ``On the rate of convergence of the levenberg-marquardt method,'' in \emph{Topics in Numerical Analysis: With Special Emphasis on Nonlinear Problems}.\hskip 1em plus 0.5em minus 0.4em\relax Springer, 2001, pp. 239--249.

\bibitem{BSY14}
S.-H. Bae and K.-J. Yoon, ``Robust online multi-object tracking based on tracklet confidence and online discriminative appearance learning,'' in \emph{Proceedings of the IEEE conference on computer vision and pattern recognition}, 2014, pp. 1218--1225.

\bibitem{BCL09}
M.~Bertogna, M.~Cirinei, and G.~Lipari, ``Schedulability analysis of global scheduling algorithms on multiprocessor platforms,'' \emph{IEEE Transactions on Parallel and Distributed Systems}, vol.~20, pp. 553--566, 2009.

\bibitem{GRS96}
L.~George, N.~Rivierre, and M.~Spuri, ``Preemptive and non-preemptive real-time uniprocessor scheduling,'' Inria, Tech. Rep., 1996.

\bibitem{LMB14}
T.-Y. Lin, M.~Maire, S.~Belongie, J.~Hays, P.~Perona, D.~Ramanan, P.~Doll{\'a}r, and C.~L. Zitnick, ``Microsoft coco: Common objects in context,'' in \emph{In Proceedings of the European Conference on Computer Vision (ECCV)}.\hskip 1em plus 0.5em minus 0.4em\relax Springer, 2014, pp. 740--755.

\bibitem{JRM18}
J.~{Cartucho}, R.~{Ventura}, and M.~{Veloso}, ``Robust object recognition through symbiotic deep learning in mobile robots,'' in \emph{2018 IEEE/RSJ International Conference on Intelligent Robots and Systems (IROS)}, 2018, pp. 2336--2341.

\bibitem{LiLa73}
C.~Liu and J.~Layland, ``Scheduling algorithms for multi-programming in a hard-real-time environment,'' \emph{Journal of the ACM}, vol.~20, no.~1, pp. 46--61, 1973.

\bibitem{HOM14}
M.~Horowitz, ``1.1 computing's energy problem (and what we can do about it),'' in \emph{2014 IEEE international solid-state circuits conference digest of technical papers (ISSCC)}.\hskip 1em plus 0.5em minus 0.4em\relax IEEE, 2014, pp. 10--14.

\end{thebibliography}

\end{document}